\documentclass[letter,scriptaddress,twocolumn, prb,showkeys]{revtex4}

	\usepackage{amsmath}
	\usepackage{makeidx}
	\usepackage{amsfonts}
	\usepackage{array}
	\usepackage[ansinew]{inputenc}
	\usepackage[usenames,dvipsnames]{pstricks}
	\usepackage{subfigure}
	\usepackage{epsfig}
	\usepackage{float}
	\usepackage{pst-grad} 
	\usepackage{pst-plot} 
	\usepackage{lipsum}
	\usepackage{graphics}

	\usepackage[colorlinks,hyperindex]{hyperref}
	\hypersetup
	{
		colorlinks,%
		citecolor=black,%
		linkcolor=black,%
		urlcolor=black,%
	}

\usepackage{siunitx}

\newcolumntype{L}[1]{>{\raggedright\let\newline\\\arraybackslash\hspace{0pt}}m{#1}}
\newcolumntype{C}[1]{>{\centering\let\newline\\\arraybackslash\hspace{0pt}}m{#1}}
\newcolumntype{R}[1]{>{\raggedleft\let\newline\\\arraybackslash\hspace{0pt}}m{#1}}

	\newcommand{\ket}[1]{\left| #1 \right\rangle}
	\newcommand{\bra}[1]{\left\langle #1 \right|}

\newcommand{\ie}{\it{i.~e.}}	
\newcommand{\eg}{\it{e.~g.}}

\makeindex

\begin{document}

	\author{Arash Khazraie}
	\email{akhazr@phas.ubc.ca}
	\author{Kateryna Foyevtsova}
	\author{Ilya Elfimov}
	\author{George A. Sawatzky}
	\affiliation{Department of Physics $\&$ Astronomy, University of British Columbia, Vancouver, British Columbia, Canada V6T 1Z1}	
	\affiliation{Stewart Blusson Quantum Matter Institute, University of British Columbia, Vancouver, British Columbia, Canada V6T 1Z4}
\title{Oxygen holes and hybridization in the bismuthates}

\begin{abstract}
Motivated by the recently renewed interest
in the superconducting bismuth perovskites,
we investigate the electronic structure
of the parent compounds $A$BiO$_{3}$ ($A=$ Sr, Ba)
using {\it ab initio} methods and tight-binding (TB) modeling.
We use the density functional theory (DFT) in the
local density approximation (LDA) to understand
the role of various interactions in
shaping the $A$BiO$_{3}$ bandstructure near the Fermi level.
It is established that interatomic hybridization
involving Bi-$6s$ and O-$2p$ orbitals plays the
most important role. Based on our DFT calculations,
we derive a minimal TB model and demonstrate that
it can describe the properties of the bandstructure
as a function of lattice distortions, such as the opening
of a charge gap with the onset of the breathing distortion
and the associated condensation of holes onto $a_{1g}$-symmetric
molecular orbitals formed by the O-$2p_{\sigma}$ orbitals
on collapsed octahedra. We also derive a single band model involving the hopping of an extended molecular orbital involving both Bi-$6s$ and a linear combination of six O-$2p$ orbitals which provides a very good description of the dispersion and band gaps of the low energy scale bands straddling the chemical potential.

\end{abstract}

\maketitle

\section{Introduction}
Hole-doped
bismuth perovskites $A$BiO$_{3}$ ($A=$ Sr, Ba)
have recently attracted a lot of attention
as one of the few examples of
transition-metal-free high-transition-temperature
oxide superconductors\cite{Cava,Sleight,Kazakov}.
The parent compounds
are no less interesting, demonstrating a variety of
temperature-driven electronic and structural
phase transitions\cite{cox,cox2}. At low temperature,
BaBiO$_{3}$ and SrBiO$_{3}$
are insulators with some characteristic
distortions from an ideal cubic
perovskite crystal structure.
Namely, oxygen octahedra around the Bi
ions exhibit alternating breathing-in and breathing-out distortions
along the three cubic crystallographic directions,
resulting in disproportionated Bi-O bond lengths.
Additionally, the O$_6$ octahedra
are tilted and rotated,
following a $a^{-}a^{-}c^{0}$ pattern in BaBiO$_{3}$
and a $a^{-}a^{-}c^{+}$ pattern in SrBiO$_{3}$
in Glazer's classification \cite{Glazer1,Glazer2}.

The insulating state of $A$BiO$_3$
is often interpreted in terms of a
charge-disproportionation model~\cite{cox,cox2,Varma,Hase}
In this model,
Bi ions with the nominal valency of $4+$
disproportionate into Bi$^{3+}$ and Bi$^{5+}$
as Ba$^{2+}_{2}$Bi$^{3+}$Bi$^{5+}$O$^{2-}_{6}$,
which produces
shorter Bi$^{5+}$--O bonds and longer Bi$^{3+}$--O bonds
and corresponds to the valence electron occupation
changing from $6s^{1}6s^{1}$ to $6s^{0}6s^{2}$ for the two Bi sites.
This scenario, however,
is not consistent with the on-site repulsion effects,
the high binding energy of the Bi-$6s$ states
as observed in Refs.~\onlinecite{Harrison}
and \onlinecite{Vielsack},
or the strongly covalent nature
of the Bi--O bonding~\cite{Harrison,Vielsack,Mattheiss}
and is not supported by spectroscopic
measurements finding little
difference in the Bi valence shell occupations
\cite{Hair,Orchard,Wertheim}.

Thanks to the weakly correlated nature
of the Bi-$6s$ and O-$2p$ electrons in $A$BiO$_3$\cite{Plumb},
an accurate theoretical description of these systems is
already possible in the framework of the density functional theory (DFT)
and local density approximation (LDA), althougth a more advanced treatment of exchange and correlation effects has been shown to result in an enhancement of the gap and the electron-phonon coupling. The states bridging the Fermi energy are basically unchanged as compared to conventional DFT, indicating that the use of LDA+$U$ or hybrid functionals merely increases the gap value.\cite{Franchini, Franchini2,Korotin,Yin,Nourafkan, Korotin2}
Previous DFT
studies\cite{Mattheiss,Mattheiss82,Mattheiss83,Foyevtsova}
generally find that
the Bi-$6s$ states lie deeper in energy by
several electron volts than the O-$2p$ states,
questioning further the charge-disproportionation model. 
In Ref.~\onlinecite{Foyevtsova}, we used
DFT methods to validate an alternative microscopic model
for the insulating state of $A$BiO$_3$,
considered initially by Ignatov\cite{Ignatov},
in which
the hole pairs condense spatially onto collapsed
O$_6$ octahedra occupying molecular
orbitals of the $a_{1g}$ symmetry
while all the Bi ions are close to being $6s^2$,
{\ie}, the following process is taking place:
\begin{align*}
2\text{Bi}^{3+}\underline{L} \to \text{Bi}^{3+}\underline{L}^{2}+\text{Bi}^{3+},
\end{align*}
where $\underline{L}$ represents a ligand hole
in an $a_{1g}$-symmetric molecular orbital on a collapsed O$_6$ octahedron.
Note that in this scenario all oxygens remain equivalent in terms of
charge and the resulting
insulating state should be named ``bond-disproportionated''
rather than ``charge-disproportionated''.

In the present follow up paper,
we intend to better understand the relevance of various interactions
in determining the electronic structure of $A$BiO$_{3}$
and derive its minimal tight-binding (TB) model. Such a minimal model describing the low energy scale states is especially useful in constructing model Hamiltonians including the electron-phonon interactions to discuss aspects such as bipolaron formation and superconductivity in hole or electron doped systems. 
We first use DFT calculations to study the hybridization strengths between the constituent
elements of $A$BiO$_{3}$, demonstrating the extreme effects of
interatomic hybridization involving
the Bi-$6s$ and O-$2p$ orbitals.
Having determined the most relevant atomic orbitals and interactions,
we then study the properties
of our derived TB model as a function of structural distortions in
two and three dimensions. Finally, we explore possible simplifications
of the TB model with a focus on describing low-energy electronic excitations.

\begin{figure*}
\begin{center}
\includegraphics[width=\textwidth]{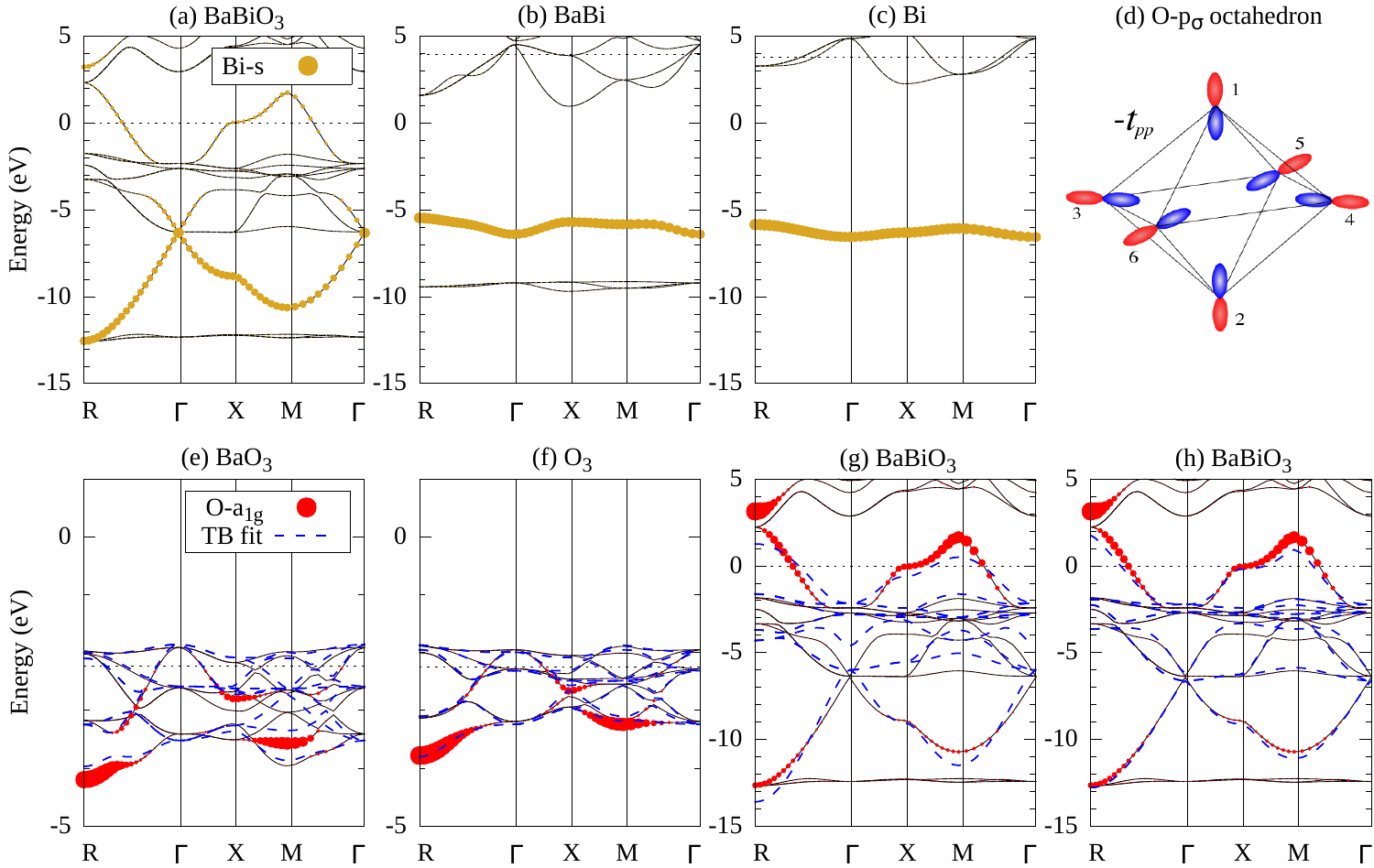} 
\caption{
The DFT (LDA) bandstructures of
(a), (g), (h) BaBiO$_{3}$ and its
(b) BaBi,
(c) Bi,
(e) BaO$_3$, and
(f) O$_3$ sublattices.
In (a) - (c),
the yellow-colored fat bands
represent the contribution of the Bi-$s$ orbital,
while in (e) - (h), the red-colored fat bands
represent the contribution of the O-$a_{1g}$ molecular orbital.
The Fermi level is marked with a
horizontal dashed black line.
Panel (d) shows the O-$a_{1g}$ molecular
orbital combination of oxygen-$p_{\sigma}$ orbitals in an
octahedron. The oxygen sites 1 to 6 are coupled via hopping integrals
$-t_{pp} = (-t_{pp\sigma}+t_{pp\pi})/2$. A nearest neighbor TB model fit of (e) BaO$_3$ (f) O$_3$ and (g)-(h) BaBiO$_3$ is shown with dashed lines. In (h), Bi-$6p$ orbitals are added in an extended tight-binding model (ETB) for an improved fit. The parameter values
resulting from the fits are listed in Table~\ref{tab:Hoppings}.}
\label{Fig:BaBiO.pdf}
\end{center}
\end{figure*}

\section{Results and discussion}
\subsection{Bi, BaBi, O$_3$, and BaO$_3$ sublattices of BaBiO$_3$}
In order to study the hybridization strengths between
the constituent elements of $A$BiO$_3$,
we calculate and compare the electronic band structures of BaBiO$_{3}$
and of its Bi, BaBi, O$_3$, and BaO$_3$ sublattices.
It is to be expected that all the conclusions
in this section will be equally applicable to
SrBiO$_3$ because of its close similarity to BaBiO$_{3}$.

The electronic structure calculations are performed
with the
full-potential linearized-augmented-planewave code
WIEN2k\cite{Blaha}.
Exchange and correlation effects
are treated within the generalized gradient density approximation
(GGA)\cite{Perdew}.
For now, we will neglect the effects of
lattice distortions
and consider
an idealized cubic unit cell containing
one formula unit.
A value of 4.34~{\AA} is used 
for the lattice constant,
taken as the average over the nearest-neighbor
Bi-Bi distances in the experimentally measured distorted
structure\cite{cox}, and a $7\times7\times7$
k-point grid is used
for the Brillouin-zone integration.

\begin{table}[t!]
\begin{tabular}{c c cccccc c cccc}
& &  \multicolumn{6}{c}{Octahedron} & & \multicolumn{4}{c}{Square plaquette} \\
\hline \hline
O site      && $a_{1g}$ & t$_{1u}$ & t$_{1u}$  & t$_{1u}$  & e$_{g}$  & e$_{g}$ & \hspace{0.0cm} & $a_{1g}$ & e$_{u}$ &  e$_{u}$ & b$_{1g}$\\
\hline
1    && $\frac{1}{\sqrt{6}}$ & $\frac{1}{\sqrt{2}}$ & 0 & 0 & $\frac{1}{\sqrt{3}}$ & 0  && $\frac{1}{\sqrt{4}}$ & $\frac{1}{\sqrt{2}}$ &0 & $\frac{1}{\sqrt{4}}$\\
2 && $\frac{1}{\sqrt{6}}$ & $\frac{-1}{\sqrt{2}}$ & 0 & 0 &  $\frac{1}{\sqrt{3}}$ & 0 && $\frac{1}{\sqrt{4}}$ & $\frac{-1}{\sqrt{2}}$ & 0 & $\frac{1}{\sqrt{4}}$\\
3   && $\frac{1}{\sqrt{6}}$ & 0 & $\frac{1}{\sqrt{2}}$ & 0 &  $\frac{-1}{\sqrt{12}}$ & $\frac{1}{\sqrt{4}}$ & & $\frac{1}{\sqrt{4}}$ & 0 &  $\frac{1}{\sqrt{2}}$  & $\frac{-1}{\sqrt{4}}$\\
4  && $\frac{1}{\sqrt{6}}$ & 0 & $\frac{-1}{\sqrt{2}}$ & 0 &  $\frac{-1}{\sqrt{12}}$ & $\frac{1}{\sqrt{4}}$ & & $\frac{1}{\sqrt{4}}$ & 0 & $\frac{-1}{\sqrt{2}}$ & $\frac{-1}{\sqrt{4}}$\\
5   && $\frac{1}{\sqrt{6}}$ & 0 & 0 & $\frac{1}{\sqrt{2}}$ &  $\frac{-1}{\sqrt{12}}$ & $\frac{-1}{\sqrt{4}}$ \\
6  && $\frac{1}{\sqrt{6}}$ & 0 & 0 & $\frac{-1}{\sqrt{2}}$ &  $\frac{-1}{\sqrt{12}}$ & $\frac{-1}{\sqrt{4}}$ \\
\hline
Energy  && $-4t_{pp}$ & 0 & 0 & 0 & $2t_{pp}$ & $2t_{pp}$ && $-2t_{pp}$ & 0 & 0 & $2t_{pp}$ \\
\hline\hline
\end{tabular} 
\caption{The eigenstates and eigenvalues
of an octahedron and a square plaquette
of O-$p_{\sigma}$ orbitals coupled via
nearest-neighbor hopping integrals
$-t_{pp}=(-t_{pp\sigma}+t_{pp\pi})/2$.
For oxygen site indexing and relative orbital phases,
refer to Fig.~\ref{Fig:BaBiO.pdf}~(d). Here, the O-$p_{\sigma}$ orbitals' on-site energies are set to zero.}
 \label{tab:3DEigen}
\end{table}

As discussed earlier\cite{Mattheiss, Foyevtsova},
the band structure
of BaBiO$_{3}$ near the Fermi level
is featured by a collection of strongly dispersive
states with predominant Bi-$6s$ and O-$2p$ orbital characters.
Most of the Bi-$6s$ character
is concentrated in the 8~eV broad lowest band
centered at $-10$~eV [Fig.~\ref{Fig:BaBiO.pdf}(a)].
In contrast,
in either the BaBi [Fig.~\ref{Fig:BaBiO.pdf}(b)] or the Bi
[Fig.~\ref{Fig:BaBiO.pdf}(c)] sublattices,
the Bi-$6s$ band width is less than 1~eV.
This indicates that interatomic hopping integrals
involving only Ba and Bi atomic orbitals are rather small.

Let us now consider
the electronic band structures of the BaO$_3$ and O$_3$ sublattices,
shown in Figs.~\ref{Fig:BaBiO.pdf}~(e) and (f),
respectively.
As we demonstrated in Ref.~\onlinecite{Foyevtsova},
it is helpful to analyze the dispersion
of the O-$2p$ states in a perovskite structure
in terms of molecular orbital combinations
of the O-$2p_{\sigma}$ atomic orbitals
in an isolated octahedron.
There are six such combinations
listed in Table~\ref{tab:3DEigen}
with the oxygen sites indexed in Fig.~\ref{Fig:BaBiO.pdf}~(d).
For future reference, Table~\ref{tab:3DEigen}
also contains molecular orbital combinations of
the O-$2p_{\sigma}$ orbitals in an isolated square
plaquette. The O-$a_{1g}$-symmetric molecular
orbital combination is of particular
interest as it is the only
combination that is allowed by symmetry
to hybridize with the Bi-$6s$ orbital.
In the calculated band structures of
BaO$_3$ and O$_3$, its character is concentrated
at the bottom of the O-$2p$ band,
mirroring the situation in an isolated octahedron
(see the bottom of Table~\ref{tab:3DEigen}).
As expected, the intensity of the O-$a_{1g}$
character is strongly k-point dependent
in the Brillouin zone of a cubic unit cell,
vanishing at the $\Gamma$-point and reaching
maximum at the $R$-point.
We find little difference
between the  widths (1.89~eV versus 2.28~eV)
and the dispersions of the O-$2p$ bands
in the BaO$_3$ and O$_3$ sublattices,
which indicates that the hybridization between the Ba and O-$2p$
orbitals is much weaker than that between the O-$2p$
orbitals themselves.
This is due to a large separation
between the O-$2p$ and Ba-$5p$ atomic
energy levels as well as a large distance of 3.04~\AA~between
the O and Ba atoms.

We can now appreciate
the enormous effect
that the Bi-$6s$ --- O-$2p$ orbital
hybridization has on the electronic structure of BaBiO$_3$,
whereby the valence band width increases from
2~eV or less in the isolated sublattices
to 15~eV in the full BaBiO$_3$ structure.
After the hybridization, the O-$a_{1g}$ molecular orbital
in an antibonding combination with the $6s$ orbital
lands above the Fermi level [see Figs.~\ref{Fig:BaBiO.pdf}~(g) or (h)].
Such behavior of the O-$a_{1g}$ molecular orbital
paves the path for the bipolaronic
condensation of oxygen holes upon breathing distortion,
as will be discussed later.

Apart from being strongly
coupled via $sp\sigma$-type overlap integrals, the hybridizing Bi-$6s$ atomic orbital
and the O-$a_{1g}$ molecular orbital
also take advantage of their energetic proximity.
To demonstrate this, the position
of the Bi-$6s$ band in, {\eg}, the Bi sublattice in Fig.~\ref{Fig:BaBiO.pdf}~(c)
has been aligned with that in BaBiO$_3$ at
the $\Gamma$ point where the Bi-$6s$ --- O-$2p$ hybridization
vanishes by symmetry, marking the Bi-$6s$ on-site energy at $\epsilon_{s}$=$-$6.1~eV. Similarly, the top
of the O-$2p$ band in the BaO$_3$ sublattice in Fig.~\ref{Fig:BaBiO.pdf}~(e)
has been aligned with the top of the O-$2p$ nonbonding states in BaBiO$_3$.
After such alignments, one clearly sees that the Bi-$6s$
orbital is only about 2~eV below the O-$a_{1g}$ molecular orbital which is much smaller than the hopping integral between the O-$a_{1g}$ and the Bi-$6s$ orbitals.

\subsection{Derivation of tight-binding models for BaO$_3$,
O$_3$, and BaBiO$_3$,}
In order to quantify the
hybridization effects discussed above, we
will now derive minimal
tight-binding models for BaBiO$_3$
and its BaO$_3$ and O$_3$ sublattices by fitting their DFT bandstructures.

In all our nearest-neighbor TB models,
there are two intersite oxygen hopping integrals
$t_{pp\sigma}$ and $t_{pp\pi}$.
We additionally include
an $sp\sigma$
hopping integral between the Ba-$6s$ and O-$2p$ 
orbitals, $t^{\text{Ba-O}}_{sp\sigma}$,
for the BaO$_{3}$ sublattice
and
an $sp\sigma$
hopping integral between the Bi-$6s$ and O-$2p$ 
orbitals, $t_{sp\sigma}$,
for BaBiO$_3$.
These simple models can nevertheless
provide an overall good description
of the DFT bandstructure, see
Figs.~\ref{Fig:BaBiO.pdf}~(e)-(g).
The parameter values
resulting from the fits are listed in Table~\ref{tab:Hoppings}.
The Bi-$6s$ --- O-$2p$ hybridization parameter $t_{sp\sigma}=2.10$~eV
is indeed found to be by far the dominant hopping integral in the system.

Surprisingly, the ratios $|t_{pp\sigma}|/|t_{pp\pi}|=5$ in O$_3$, and $|t_{pp\sigma}|/|t_{pp\pi}|=10$ in BaBiO$_3$
are considerably larger than
the empirical ratio of 3 typically assumed in cuprates \cite{Harrison2, McMahan}.
The origin of such an enhancement of the
$|t_{pp\sigma}|/|t_{pp\pi}|$ ratio is not clear, although
we have ruled out a possible sensitivity
of this parameter to the system's dimensionality and to the O-O bond length
variation by comparing calculations with accordingly modified structural parameters.

\begin{table}[tb!]
 \begin{tabular}{l S[table-format=3.2] S[table-format=3.2]  S[table-format=3.2] S[table-format=3.2]  S[table-format=3.2] }

 & \multicolumn{1}{C{1.5cm}}{O$_3$} & \multicolumn{1}{C{1.5cm}}{BaO$_3$} & \multicolumn{3}{c}{BaBiO$_3$}  \\
\hline\hline
     & TB & TB & \multicolumn{1}{C{1.0cm}}{TB} & \multicolumn{1}{C{1.0cm}}{WF} & \multicolumn{1}{C{1.0cm}}{ETB} \\
\hline
$\epsilon_{s}$                &     &      & -4.73   & -6.93  & -6.2  \\
$\epsilon_{p_{\sigma}} $      &-2.57&-2.92 &  -5.13  & -4.84  & -3.11   \\
$\epsilon_{p_{\pi}} $         &-2.49& -2.82& -3.06   &  -3.73 & -2.78  \\
$t_{pp\sigma}$                &0.26 & 0.30 & 0.63    &  0.64  &  0.4 \\
$t_{pp\pi}$                   &-0.05& -0.01& -0.04   &  -0.03 &  -0.03  \\
$t_{sp\sigma}$                &     &      &  2.10   & 2.31   &   2.09 \\
\\
$\epsilon_s^{\text{Ba}}$     & & 3.47 \\
$t_{sp\sigma}^{\text{Ba-O}}$ & & 0.9 \\
\\
$t'_{pp\sigma}$               &&  &   &   -1.0         &     \\
\\
$\epsilon_p^{\text{Bi}}$      &&  &          &             &  2.1 \\
$t_{pp\sigma}^{\text{Bi-O}}$  &&  &          &             &  2.34  \\
$t_{pp\pi}^{\text{Bi-O}}$     &&  &          &             &  -0.53 \\
\hline\hline
\end{tabular} 
\caption{On-site energies and hopping
integrals in eV for
BaBiO$_3$ and its O$_3$ and BaO$_3$ sublattices.
The values are obtained
by fitting either the simplest or the extended
tight-binding (TB or ETB) model
or by using Wannier functions (WF) including Bi-$6s$ and O-$2p$ orbitals. This choice of WF orbitals results in a large next-nearest-neighbor hopping integral $t'_{pp\sigma}$ between the O-$p_{\sigma}$ orbitals.}
 \label{tab:Hoppings}
\end{table}

\begin{figure*}
\begin{center}
\includegraphics[width=\textwidth]{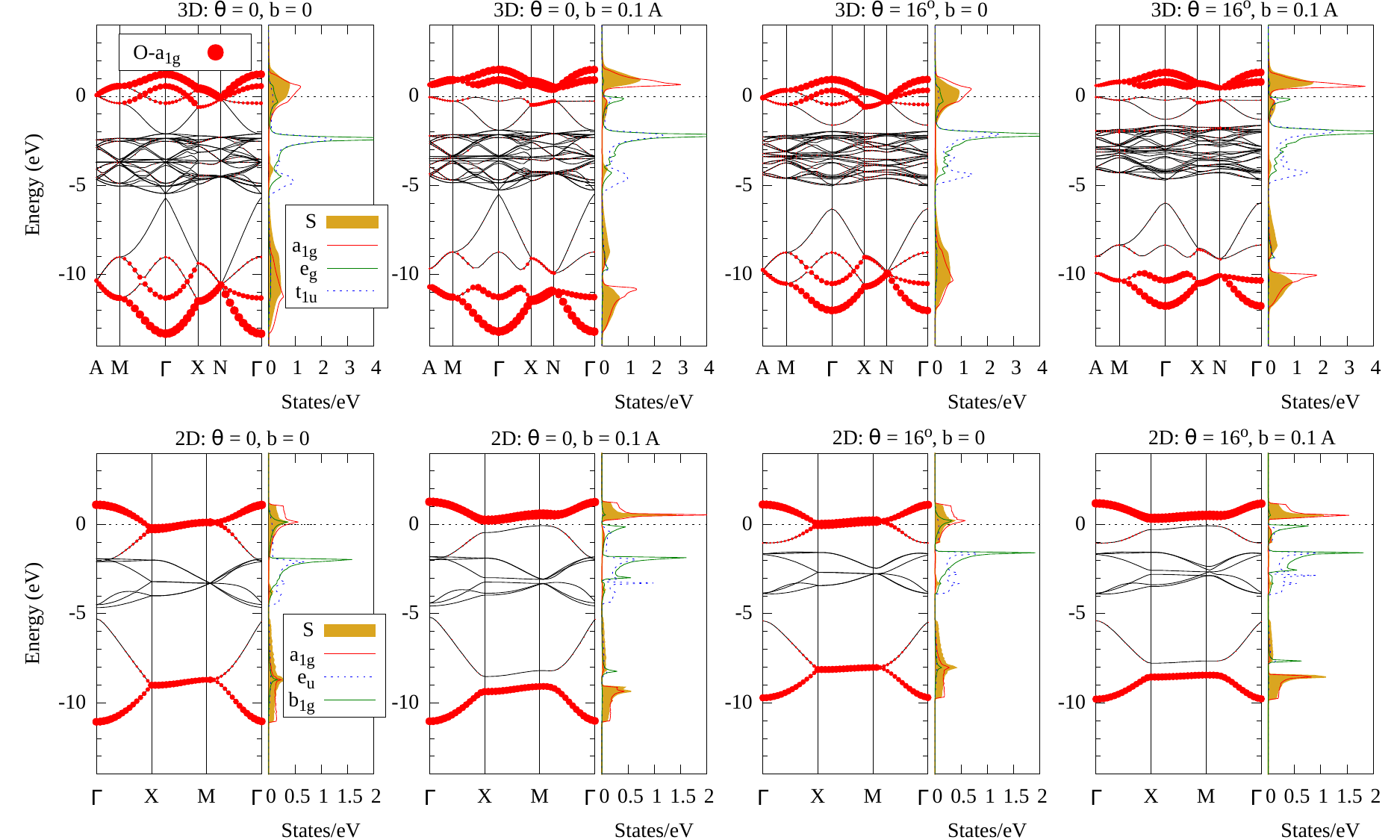} 
\caption{
The bandstructures and projected DOS
of the 3D (top panels) and 2D (bottom panels) TB models with varying strengths of
the breathing and tilting distortions. Here, breathing $b$ is half the difference between
the two disproportionated Bi-O bond lengths,
and $\theta$ is the tilting angle.
Molecular orbital projections are made for
the compressed octahedron or square plaquette
following Table~\ref{tab:3DEigen},
while the Bi-$s$ orbital projection is made for the
Bi atom located inside the compressed octahedron or square plaquette.
The red-colored fat bands represent the contribution of the
O-$a_{1g}$ molecular orbital.}
\label{Fig:3DTB.pdf}
\end{center}
\end{figure*}

For BaBiO$_3$, we also find that the TB model
parameter values
are in good agreement with hopping
integrals calculated using Wannier function (WF) projections\cite{Mostofi08,Kunes10}, where we only included Bi-$6s$ and O-$2p$ orbitals, since the low energy states spanning the Fermi energy are primarily of O-$2p$ and Bi-$6s$ character (see the fourth column of Table~\ref{tab:Hoppings}).
However, this technique gives
an unphysically large next-nearest-neighbor
hopping between the O-$p_{\sigma}$ orbitals $t'_{pp\sigma}$.
This result is a consequence of the rather strong hybridization of the O-$2p$ orbitals
with the empty Bi-$6p$
orbitals, which have not been included
in the Wannier basis.
It has motivated us to also consider
an extended tight-binding (ETB) model
with added hybridization
between the O-$2p$ and Bi-$6p$ orbitals (see the fifth column of Table~\ref{tab:Hoppings}).
The ETB model indeed provides an improved
description of the DFT bandstructure [see Fig.~\ref{Fig:BaBiO.pdf}~(h)],
but also gives a more realistic value for
the Bi-$6s$ orbital on-site energy $\epsilon_s=-6.2$~eV,
which is very close to the value of $-6.1$~eV
corresponding to the position of the
strongest Bi-$6s$ character band at $\Gamma$ in Fig.~\ref{Fig:BaBiO.pdf}~(a).

\subsection{Breathing and tilting distortions in two and
three dimensions}
In this section, we will
consider lattice distortions present in the real $A$BiO$_3$ structure.
We are interested in whether
our TB model
can capture the changes in the electronic structure
due to the lattice distortions as observed in DFT
calculations\cite{Foyevtsova}, such as the opening of the charge gap with
the on-set of the breathing distortion.
Behaviors of the TB model in three and two dimensions
will be compared to study the role of dimensionality in the problem.

Since our focus is mainly on
the top valence band crossing the Fermi level,
which is of the O-$a_{1g}$
symmetry and does not mix with
the Bi-$6p$ orbital,
we will use here the simpler TB model
from the third column of Table~\ref{tab:Hoppings}
with only Bi-$6s$ and O-$2p$
orbitals in the basis.
The coupling of electrons to lattice distortions
is modeled through a $1/d^2$
dependence of hopping integrals on the
interatomic separation $d$\cite{Harrison2}.

In order to study
the individual roles of the breathing and tilting
distortions, let us consider four model
structures of $A$BiO$_3$ with the following
characteristics:
(i) $b=0$~{\AA}, $\theta=0${\textdegree},
(ii) $b=0.1$~{\AA}, $\theta=0${\textdegree},
(iii) $b=0$~{\AA}, $\theta=16.5${\textdegree}, and
(iv) $b=0.1$~{\AA}, $\theta=16.5${\textdegree}.
Here, $b$ is half the difference between
the two disproportionated Bi-O bond lengths
and $\theta$ is the tilting angle of the octahedron in three-dimensions, or the rotation of the square plaquette in two-dimensions.
The values of $b=0.1$~{\AA} and $\theta=16.5${\textdegree} in structure (iv)
correspond to the respective strengths of the breathing and tilting distortions
in the experimental SrBiO$_3$ structure.
Because the distortions break translational symmetry,
there are four formula units in
the three-dimensional (3D) unit cell
and two formula units in the two-dimensional (2D) unit cell.

\begin{table}
\begin{tabular}{l*{7}{C{1.85cm}}}
\hline\hline
      & $t_{sp{\sigma}}$~(eV)
      & $t_{pp\sigma}$~(eV)
      & $t_{pp\pi}$~(eV) \\
\hline
$b=0$~{\AA},   $\theta=0${\textdegree} & 2.1    &  0.63 &   -0.04 \\
$b=0.1$~{\AA}, $\theta=0${\textdegree}
& 2.37  \hspace{0.05cm} 1.96
& 0.71  \hspace{0.05cm} 0.59 
& -0.045 \hspace{0.05cm} -0.03\\
\hline\hline
\end{tabular} 
\caption{Variation of the
nearest-neighbor hopping integrals
in response to the Bi-O bond-disproportionation of 0.1 \AA.}

 \label{tab:hopping}
\end{table}
Figure~\ref{Fig:3DTB.pdf} presents the bandstructures and
the projected densities of states (DOS)
of our 3D (top panels) and 2D (bottom panels) model structures. Molecular orbital projections are made for the compressed octahedron or square plaquette
following Table~\ref{tab:3DEigen},
and the red-colored fat bands represent the contribution of the
O-$a_{1g}$ molecular orbital.
We find that the models' electronic structures
exhibit similar characteristics
irrespective of the dimensionality.
Close to the Fermi level,
the tilting distortion
opens a gap at around -1.5~eV
and causes an overall band narrowing
while the breathing distortion opens
a charge gap transforming the system into a semiconductor.
This metal-to-semiconductor
transition is accompanied by a shift
of the O-$a_{1g}$ molecular orbital character
into the empty states. Its intensity becomes
k-point independent meaning
that in real space holes spatially condense into
well-defined molecular orbitals on the collapsed
octahedra.
The observed strong tendency towards formation of
molecular orbitals can be due to the fact
that oxygen hopping integrals are rather sensitive
to the Bi-O bond-disproportionation.
As one can see in Table~\ref{tab:hopping},
bond-disproportionation of 0.1~{\AA}~results in 0.4 eV difference in the $t_{sp\sigma}$ hopping integrals for the collapsed and the expanded octahedron.

\begin{figure}[t]
\begin{center}

\includegraphics[width=0.48\textwidth]{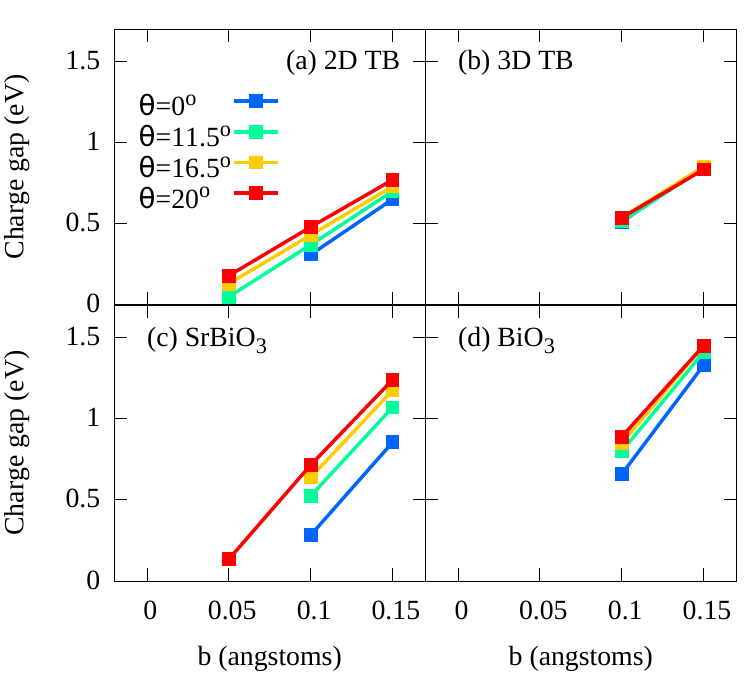} 
\caption{The charge gap as a function of the breathing
distortion at various tilting distortions
in 
(a) the 2D TB model,
(b) the 3D TB model,
(c) SrBiO$_3$ from DFT (LDA) calculations\cite{Foyevtsova}, and
(d) the BiO$_3$ sublattice
from DFT (LDA) calculations.}
\label{Fig:gap}
\end{center}
\end{figure}
Let us also have a closer look at the behavior
of the charge gap as a function of $t$ and $b$.
It is depicted in Figs.~\ref{Fig:gap}~(a) and (b)
for the 2D and 3D TB models, respectively.
In both cases, the gap is linear in $b$
but its $\theta$-dependence is stronger in the 2D case.
Qualitatively our model calculations
can reproduce the
DFT results for SrBiO$_3$\cite{Foyevtsova}
[Fig.~\ref{Fig:gap}~(c)], but quantitatively
the effects of both the breathing and tilting distortions
are rather underestimated, especially in the 3D case.
This is due to the approximations we used in the
model calculations, such as neglecting the variation of
on-site energies
or limiting the number of orbitals.
To exemplify the effect of the latter approximation,
we compare the DFT gap in SrBiO$_3$
with that in BiO$_3$ [Fig.~\ref{Fig:gap}~(d)]. Here,
the $\theta$-dependence of the gap is noticeably reduced,
similarly to what we find in the model calculations
where the $A$ cation orbitals are also neglected.
This suggests that
hybridization with the $A$ cation orbitals
plays a role in determining the size of the charge gap as a function of $\theta$.
 \begin{figure*}[t]
\begin{center}
\includegraphics[width=0.99\textwidth]{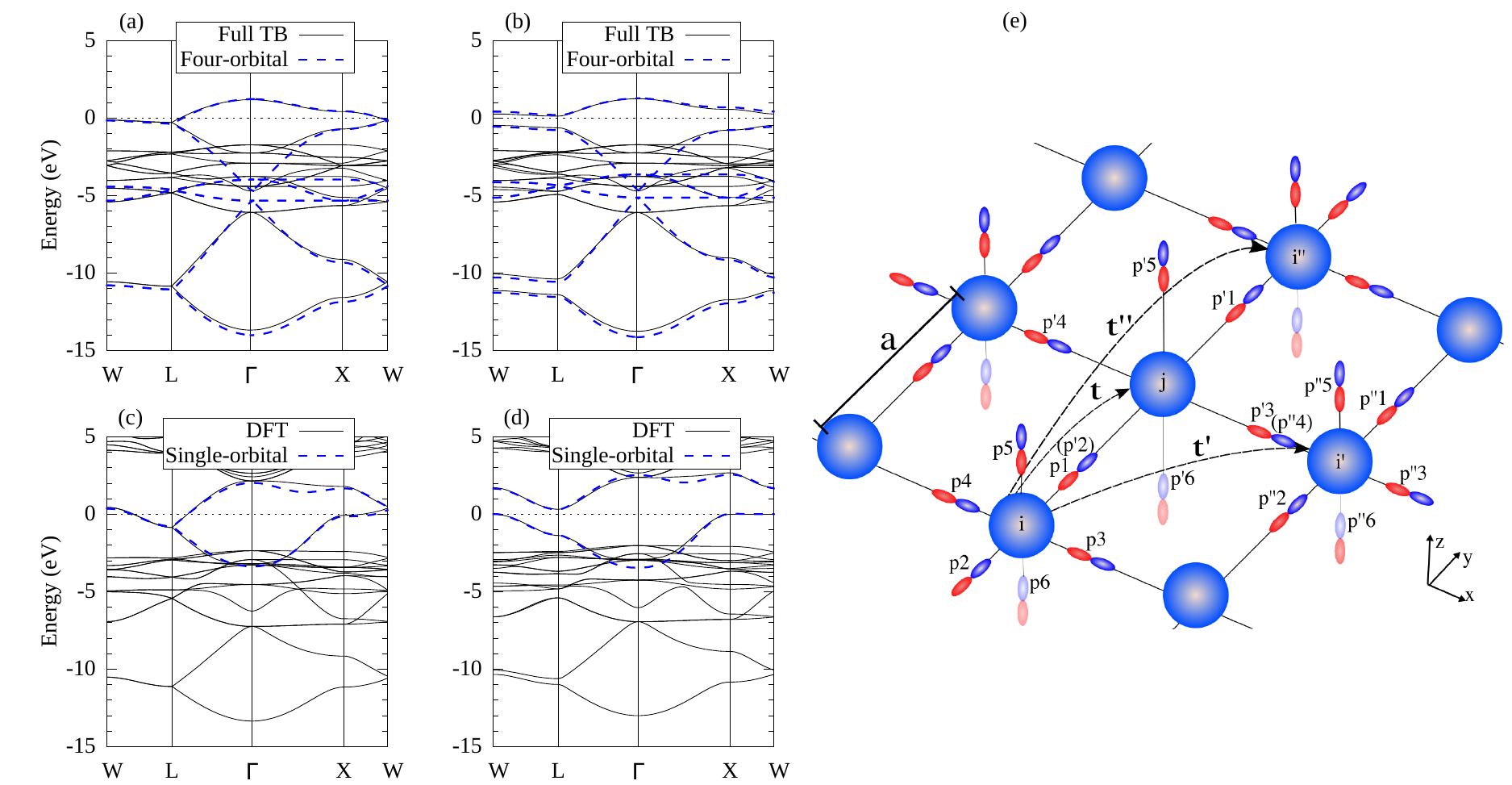} 
\caption{In (a) and (b),
the full TB (solid line) and the four-orbital TB (dashed line) models are compared
for the $b=0.0$~{\AA} and $b=0.1$~{\AA} lattices,
respectively.
In (c) and (d), the DFT bandstructure  (solid line) and
the single-orbital TB model (dashed line) are compared
for the $b=0.0$~{\AA} and $b=0.1$~{\AA} lattices,
respectively. In (e), the single-orbital $A_{1g}$ coupling to nearest, second nearest, and fourth nearest neighbors are shown.}
\label{Fig:simplemodel}
\end{center}
\end{figure*}

\subsection{Tight-binding models with a reduced number of orbitals}
Finally, we discuss possible
simplifications of the $A$BiO$_3$ TB model
that would still allow an accurate description
of low-energy electronic excitations.
As was shown previously,
the bands straddling the Fermi level
are dominantly of the Bi-$s$ and O-$p_{\sigma}$ orbital character. Therefore, a natural simplification of the TB model could be to eliminate the O-$p_{\pi}$ orbitals from the basis. This reduces the basis size
from ten to four orbitals per formula unit.
As one can see in Figs.~\ref{Fig:simplemodel}~(a)
and (b), the four-orbital TB model gives
a good agreement with the full ten-orbital model
near the Fermi level
even without adjustment of model parameters.
Here, the calculations are done for a
face-centered cubic unit cell with two Bi sites,
and the non-distorted lattice [panel (a)]
is compared with a lattice featuring an 0.1~{\AA}
breathing distortion and no octahedra tilting [panel (b)].
The comparison illustrates, in particular,
that the four-orbital model is capable of
describing the distortion-induced
metal-to-semiconductor transition in $A$BiO$_3$.

Despite its reduced basis size,
the four-orbital model, however, contains
redundant degrees of freedom,
as far as low-energy physics is concerned.
They give rise to the bonding Bi-$s$ and O-$p_{\sigma}$
states at $-$10~eV and the oxygen nonbonding
states at $-$3~eV, {\ie} in the energy regions
deep below the Fermi level.
One can take a step further and
write down a single-orbital TB model
with an $A_{1g}$-symmetric orbital at each octahedron site. For this model, which could represent only the low energy scale bands, the basis consists of antibonding combinations of Bi-$s$ and O-$a_{1g}$ orbitals:
\begin{align*}
 \ket{\psi^{A_{1g}}} = \frac{1}{\sqrt{\alpha^{2}+\beta^{2}}}\Big( \alpha\ket{\psi^{\text{Bi}-s}}-\beta\ket{\psi^{\text{O}-a_{1g}}}\Big)
\end{align*}
where $\psi^{\text{O}-a_{1g}}$ orbital is a symmetric linear combination of O-p$_{\sigma}$[Fig.~\ref{Fig:BaBiO.pdf}~(d)].
Neglecting spin, the effective Hamiltonian in this basis can be written as:
\begin{align*}
&
H = \sum_{i} \epsilon^{c}_{A_{1g}}\hat{c}^{\dagger}_{i}\hat{c}_{i} + \sum_{j} \epsilon^{e}_{A_{1g}}\hat{d}^{\dagger}_{j}\hat{d}_{j}  + H^{c-e} + H^{c-c} + H^{e-e} .
\end{align*}
Here, indices $i$ and $j$ run over collapsed and expanded octahedron sites, respectively; $\hat{c}^{\dagger}_{i}$ ($\hat{c}_{i}$) create (annihilate) a hole on site $i$ and $\hat{d}^{\dagger}_{j}$ ($\hat{d}_{j}$) create (annihilate) a hole on site $j$; $\epsilon^{c}_{A_{1g}}$ ($\epsilon^{e}_{A_{1g}}$) is the on-site energy of the $A_{1g}$  orbital on a collapsed (expanded) octahedron site.
The hybridization terms can be written as:
\begin{align*}
&
H^{c-e} = \sum^{n.n.}_{<ij>} t\hat{c}^{\dagger}_{i}\hat{d}_{j}  + h.c.\\
&
H^{c-c} = \sum_{i}  \sum_{i' \in \{i\}}t'\hat{c}^{\dagger}_{i}\hat{c}_{i'} +   \sum_{i}\sum_{i'' \in \{i\}}t''\hat{c}^{\dagger}_{i}\hat{c}_{i''} + h.c.\\
&
H^{e-e} = \sum_{j}\sum_{j' \in \{j\}}t'\hat{d}^{\dagger}_{j}\hat{d}_{j'} +   \sum_{j}\sum_{j'' \in \{j\}}t''\hat{d}^{\dagger}_{j}\hat{d}_{j''}  + h.c.
\end{align*}
where $<ij>$ represents sum over nearest-neighbor sites, $i'$ and $i''$ ($j'$ and $j''$) are sites at distances $\sqrt{2}a$ and $2a$ from site $i (j)$, respectively [Fig.~\ref{Fig:simplemodel}(e)]. The model parameter values are obtained by fitting to the DFT states closest to the Fermi level. 
The parameter values are given in Table~\ref{tab:singleorb}
for lattices with a varying degree of the breathing distortion,
and the fits for the $b=0$~{\AA} and $b=0.1$~{\AA} lattices
are shown in Figs.~\ref{Fig:simplemodel}~(c) and (d), respectively.
\begin{table}[t]
\begin{tabular}{l S[table-format=3.2] S[table-format=3.2]  S[table-format=3.2] S[table-format=3.2]  S[table-format=3.2] S[table-format=3.2]}
\hline\hline
                \multicolumn{1}{C{1.0cm}}{$b$}
              & \multicolumn{1}{C{1.0cm}}{$\epsilon^{c}_{A_{1g}}$}
              & \multicolumn{1}{C{1.0cm}}{$\epsilon^{e}_{A_{1g}}$}
              & \multicolumn{1}{C{1.0cm}}{$t$}
              & \multicolumn{1}{C{1.0cm}}{$t'$}
              & \multicolumn{1}{C{1.0cm}}{$t''$} 
              & \multicolumn{1}{C{1.5cm}}{B.D.} \\
\hline
$0.00$~{\AA}&-0.13&-0.13 & -0.45   & -0.09  &  0.10 & 1.00 \\
$0.05$~{\AA}&0.35&-0.51 & -0.47   & -0.10  &  0.11  & 1.46 \\
$0.10$~{\AA}& 0.99& -0.65& -0.48   & -0.11  &  0.115 & 1.68 \\
$0.15$~{\AA}&1.86 & -0.78 & -0.50   & -0.12  &  0.125 & 1.78\\
\hline\hline
\end{tabular} 
\caption{The single-orbital TB model parameter values in eV
for lattices with a varying degree of the breathing distortion $b$
and no octahedra tilting.
$t$, $t'$, and $t''$ are the nearest, second-nearest,
and fourth-nearest neighbor hopping integrals, respectively.
$\epsilon^{c}_{A_{1g}}$ and $\epsilon^{e}_{A_{1g}}$
are the on-site energies of the $A_{1g}$-like orbitals
of the collapsed and expanded octahedron. Bond disproportionation (B.D.) shows the number of $A_{1g}^{c}$ orbital holes. }
\label{tab:singleorb}
\end{table}
We find that within the single-orbital TB model
the appearance and growth of the charge gap
with an increasing breathing distortion
can be well described by a splitting
of the two $A_{1g}$ orbital on-site energies, with essentially no
need of modifying the hopping integrals [see Table~\ref{tab:singleorb}].
This model can be interpreted as an effective low energy model of bismutathes that can well describe the bands near the Fermi level and can now be used for example to include electron-phonon coupling keeping in mind the origin of these wave functions and the effects of electron or hole doping looking for possible superconductivity. 

 In order to clarify the physics involved in these effective hopping integrals, it is instructive to obtain estimates for hopping integrals, $t$, $t'$, and $t''$ by considering the composition of the $A_{1g}$ orbitals described above. These hoppings can be directly related to our full ten orbital model parameters, $t_{pp\sigma}$, $t_{pp\pi}$, and $t_{sp\sigma}$, with taking into account O-O and Bi-O hoppings only up to nearest-neighbor. In the following, we consider for simplicity a non-distorted case, approximate the coefficients $\alpha$ and $\beta$ to be $\alpha=\beta=1$, and also neglect the nonorthogonality that occurs in the O-$a_{1g}$ orbitals on nearest-neighbors in estimating $t$, since the overlap is only $\frac{1}{6}$.
The $A_{1g}$ nearest-neighbor hopping $t$ [see Fig.~\ref{Fig:simplemodel}(e)] can be written as:

\begin{widetext}
\begin{align*}
&
t=\bra{\psi^{A_{1g}}_{i}}H\ket {\psi^{A_{1g}}_{j}}=\\
&
\bra{\frac{1}{\sqrt{2}}\Big(\psi^{s}-\frac{1}{\sqrt{6}}(-p_{1}+p_{2}-p_{3}+p_{4}-p_{5}+p_{6})\Big)_{i}}H
\ket{\frac{1}{\sqrt{2}}\Big(\psi^{s}-\frac{1}{\sqrt{6}}(-p'_{1}+p'_{2}-p'_{3}+p'_{4}-p'_{5}+p'_{6})\Big)_{j}}\\
&
=-\frac{1}{2\sqrt{6}}\bra{\psi^{s}_{i}}H\ket{(p'_{2})_{j}}+\frac{1}{2\sqrt{6}}\bra{p_{1_{i}}}H\ket{\psi^{s}_{j}}+\frac{1}{12}\bra{-p_{1_{i}}}H\ket{(-p'_{3}+p'_{4}-p'_{5}+p'_{6})_{j}}\\
&
+\frac{1}{12}\bra{(-p_{3}+p_{4}-p_{5}+p_{6})_{i}}H\ket{(p'_{2})_{j}} = -\frac{1}{\sqrt{6}}t_{sp\sigma}+\frac{2}{3}t_{pp} = -\frac{1}{\sqrt{6}}t_{sp\sigma}+\frac{1}{3}(t_{pp\sigma}-t_{pp\pi}) \approx -0.64~\text{eV},
\end{align*}
\end{widetext}
The next-nearest neighbor hopping term can be written:
\begin{widetext}
\begin{align*}
&
t'=\bra{\psi^{A_{1g}}_{i}}H\ket {\psi^{A_{1g}}_{i'}}=\\
&
\bra{\frac{1}{\sqrt{2}}\Big(\psi^{s}-\frac{1}{\sqrt{6}}(-p_{1}+p_{2}-p_{3}+p_{4}-p_{5}+p_{6})\Big)_{i}}H\ket{\frac{1}{\sqrt{2}}\Big(\psi^{s}-\frac{1}{\sqrt{6}}(-p''_{1}+p''_{2}-p''_{3}+p''_{4}-p''_{5}+p''_{6})\Big)_{ i'}}\\
&
=\frac{1}{2}\times\frac{1}{6}\bra{(p_{1} + p_{3})_{i}}H\ket{(-p''_{2}-p''_{4})_{i'}}= -\frac{2}{12}t_{pp} =\frac{1}{12}(-t_{pp\sigma}+t_{pp\pi}) \approx -0.05~\text{eV},
\end{align*}
\end{widetext}

and the fourth nearest-neighbor term:
\begin{widetext}
\begin{align*}
&
t''=\bra{\psi^{A_{1g}}_{i}}H\ket {\psi^{A_{1g}}_{i''}}\\
&
=\bra{\psi^{A_{1g}}_{i}}\psi^{A_{1g}}_{j}\Big \rangle \bra{\psi^{A_{1g}}_{j}}H\ket {\psi^{A_{1g}}_{i''}}=\bra{\psi^{A_{1g}}_{i}}\psi^{A_{1g}}_{j}\Big \rangle \times t\\
&
=\bra{\frac{1}{\sqrt{2}}\Big(\psi^{s}-\frac{1}{\sqrt{6}}(-p_{1}+p_{2}-p_{3}+p_{4}-p_{5}+p_{6})\Big)_{i}}\frac{1}{\sqrt{2}}\Big(\psi^{s}-\frac{1}{\sqrt{6}}(-p'_{1}+p'_{2}-p'_{3}+p'_{4}-p'_{5}+p'_{6})\Big)_{j}\bigg\rangle \times t\\
&
= \frac{1}{2}\times\frac{1}{6}\bra{-p_{1}}p'_{2}\rangle\times t \approx +0.05~\text{eV}
\end{align*}
\end{widetext}
where $\bra{\psi^{A_{1g}}_{i}}\psi^{A_{1g}}_{j}\Big\rangle=-\frac{1}{12}$ is the overlap integral between site $i$ and $j$. The above estimates are within the order of magnitude of the single-orbital parameters in Table~\ref{tab:singleorb} obtained from the fit.
For the on-site energies we have:
\begin{align*}
&
\epsilon_{A_{1g}} =\bra{\psi^{A_{1g}}_{i}}H\ket {\psi^{A_{1g}}_{i}}\\
&
=\frac{1}{2}\bra{\Big(\psi^{s}-\psi^{a_{1g}}\Big)_{i}}H\ket {\Big(\psi^{s}-\psi^{a_{1g}}\Big)_{i}}\\
&
=\frac{1}{2}\Big(\epsilon_{s}+\epsilon_{a_{1g}}\Big)-\frac{6}{\sqrt{6}}t_{sp\sigma}
\end{align*}
where on-site energy of O-$a_{1g}$ is $\epsilon_{a_{1g}}=\epsilon_{\sigma}-4t_{pp}$ and $\frac{6}{\sqrt{6}}t_{sp\sigma}$ is the coupling energy of Bi-$s$ and O-$a_{1g}$ orbitals.
The change in the on-site energies due to breathing can now be written as:
\begin{align*}
\Delta \epsilon = \epsilon^{c}_{A_{1g}} - \epsilon^{e}_{A_{1g}} = (-2t^{c}_{pp} - \frac{6}{\sqrt{6}}t^{c}_{sp\sigma})-(-2t^{e}_{pp} - \frac{6}{\sqrt{6}}t^{e}_{sp\sigma})\\.
\end{align*}
Here, t$^{c}_{pp}(t^{e}_{pp}$) represent O-O hoppings [see Fig.\ref{Fig:BaBiO.pdf}(d)] of the collapsed (expanded) $A_{1g}$ orbitals, and $t^{c}_{sp\sigma}(t^{e}_{sp\sigma})$ are the corresponding Bi-O hoppings, and we have assumed that Bi-$s$ on-site energies are unchanged. Using hopping parameters of a distorted lattice of 0.1 \AA~from Table~\ref{tab:hopping}, a direct gap value of $ \Delta \epsilon \approx 1.1$ eV is estimated. This value of the gap is comparable to $\Delta \epsilon$ in Table~\ref{tab:singleorb} obtained from the fit.

In Table~\ref{tab:singleorb} we show an effective orbital occupation corresponding to the number of holes in the collapsed $A_{1g}^{c}$ orbital calculated by integrating the projected density of states above the chemical potential as a function of the breathing distortion. Since we have two Bi and two holes per unit cell the number of holes in the expanded $A_{1g}^{e}$ orbital is also 1.0 for no breathing. As the distortion increases we see that the number of holes in the collapsed $A_{1g}^{c}$ molecular orbital gradually moves towards 2.0 at which point there would be no holes anymore in the expanded $A_{1g}^{e}$ molecular orbital. This looks very much like charge-disproportionation. However, in the structure all the oxygens are identical which means that this cannot be charge-disproportionation involving the oxygen. Each oxygen indeed participates in both the collapsed and expanded $A_{1g}$ molecular orbitals however it participates more in the collapsed $A_{1g}^{c}$ molecular orbital than in the expanded one. So again, this has to do with bond-disproportionation. We note however that there is no symmetry change in moving from a bond to a charge-disproportionation picture and so there is no clear boundary but rather a gradual cross-over. A real charge-disproportionation would have to imply an attractive coulomb interaction while a bond-disproportionation results from the electron density changes in the bonds driven by an electron-phonon coupling involving the hopping integral changes.

We can infer the electron-phonon coupling strength from our single-orbital model as the $A_{1g}$ molecular orbital on-site energy lowering in Table~\ref{tab:singleorb}, which is about 1 eV in the presence of the experimental bond modulation of 0.1 \AA~or equivalently $\frac{d\epsilon}{dx}=$ 10 eV/\AA. Given a Raman breathing mode phonon frequency of $\omega_{ph} \approx$ 70~meV\cite{Sugai,Tajima} at $q=(\pi,\pi,\pi)$ for BaBiO$_{3}$ the electron-phonon coupling $g$ can be estimated as:
\begin{align*}
g = \frac{\partial \epsilon}{\partial x}\sqrt{\frac{\hbar}{2M_{o}\omega_{ph}}}\approx 10~\text{eV/\AA} \times 0.04321~\text{\AA}\approx 0.4321~\text{eV}.
\end{align*}
where $M_{o}$ is the oxygen mass.
The electron-phonon coupling $g$ translated into the dimensionless coupling $\lambda$ is:
\begin{align*}
\lambda=\frac{2g^{2}}{\hbar\omega_{ph} W}=\frac{2\times0.4321^{2}}{0.07\times 6}\approx 0.89,
\end{align*}
where $W\approx 6$ eV is the single-orbital $A_{1g}$ bandwidth read from Fig.~\ref{Fig:simplemodel}~(d). Following the same argument we can estimate a value of $\lambda =$ 1.21  for Ba$_{0.6}$K$_{0.4}$BiO$_{3}$ given a breathing mode phonon frequency of $\omega_{ph} \approx$ 60~meV\cite{Braden} for Ba$_{0.6}$K$_{0.4}$BiO$_{3}$ and assuming that the bandwidth $W$ is unchanged upon K substitution. These electron-phonon couplings are much higher than what has been obtained from previous LDA calculations on Ba$_{0.6}$K$_{0.4}$BiO$_{3}$\cite{Hamada, Shirai, Liechtenstein, Kunc, Meregalli, Bazhirov}. Of course, to make a real comparison with LDA estimates one would have to determine the $q$ dependence of this coupling and average this over an assumed Fermi surface. 

\section{Conclusions}
In summary, we have studied
the electronic structure of the bismuth perovskites $A$BiO$_3$ ($A$ = Sr, Ba)
using {\it ab initio} calculations and tight-binding modeling.
We found that the hopping integrals
involving the Bi-$6s$ and O-$2p$ orbitals
play a leading role in shaping
the electronic bandstructure of $A$BiO$_3$ near the Fermi level.
A minimal TB model with ten orbitals per formula unit (one Bi-$6s$
and nine O-$2p$ orbitals) was derived and shown
to be able to describe the changes in the electronic structure
due to lattice distortions that had been
observed in previous DFT studies, such as
the opening of the charge gap due to the breathing distortion
and the associated formation of molecular orbitals on
collapsed octahedra.
We also showed that for the purpose of exploring
low-energy excitations in $A$BiO$_3$ this TB model
can be further reduced to a four-orbital
one with one Bi-$6s$
and three O-$2p_{\sigma}$ orbitals in the basis
and even further down to a single-orbital one with a single $A_{1g}$-like orbital at each octahedron site. Within this model, we further estimated electron-phonon couplings of $\lambda =$ 0.89
and $\lambda =$ 1.21 for BaBiO$_{3}$ and Ba$_{0.6}$K$_{0.4}$BiO$_{3}$ respectively.
This single band model is a good representation of the band structure close to the chemical potential and can be used in a more detailed study including the influence of electron-phonon coupling and possible mechanisms for superconductivity in the doped materials, but one has to keep in mind the rather extended molecular orbital character of the basis states in such a model.

This work was supported by NSERC, CIFAR, and the Max Planck-UBC Stewart Blusson Quantum Matter Institute. 


\begin{thebibliography}{}
 \bibitem{Cava}
R. J. Cava, B. Batlogg, J. J. Krajewski, R. Farrow, L. W. R. Jr., A. E. White, K. Short, W. F. Peck, and T. Kometani, Nature (London) \textbf{332}, 814 (1988).

 \bibitem{Sleight}
A. Sleight, J. Gillson, and P. Bierstedt, Solid State Communications \textbf{17}, 27 (1975).

 \bibitem{Kazakov}
S. M. Kazakov, C. Chaillout, P. Bordet, J. J. Capponi, M. Nunez-Regueiro, A. Rysak, J. L. Tholence, P. G. Radaelli, S. N. Putilin, and E. V. Antipov, Nature (London) \textbf{390}, 148 (1997).

 \bibitem{cox}
  D. Cox and A. W. Sleight, Solid State Communications \textbf{19}, 969 (1976).

   \bibitem{cox2}
  D. E. Cox and A. W. Sleight, Acta Crystallographica Section B  \textbf{35}, 1 (1979).

   \bibitem{Glazer1}
  A. M. Glazer, Acta Crystallographica Section B  \textbf{28}, 3384 (1972).

    \bibitem{Glazer2}
  A. M. Glazer, Acta Crystallographica Section A  \textbf{31}, 756 (1975).

     \bibitem{Varma}
  C. M. Varma, Phys. Rev. Lett. \textbf{61}, 2713 (1988).

     \bibitem{Hase}
  I. Hase and T. Yanagisawa, Phys. Rev. B \textbf{76}, 174103 (2007).

       \bibitem{Harrison}
  W. A. Harrison, Phys. Rev. B \textbf{74}, 245128 (2006).

        \bibitem{Vielsack}
  G. Vielsack and W. Weber, Phys. Rev. B \textbf{54}, 6614 (1996).

       \bibitem{Mattheiss}
  L. F. Mattheiss and D. R. Hamann, Phys. Rev. B \textbf{28}, 4227 (1983).

 \bibitem{Hair}
  J. de Hair and G. Blasse, Solid State Communications \textbf{12}, 727 (1973).
  
         \bibitem{Orchard}
  A. F. Orchard and G. Thornton, J. Chem. Soc. Dalton Trans. , 1238 (1977).

         \bibitem{Wertheim}
  G. K. Wertheim, J. P. Remeika, and D. N. E. Buchanan, Phys. Rev. B \textbf{26}, 2120 (1982).

 \bibitem{Plumb}
N. C. Plumb, D. J. Gawryluk, Y. Wang, Z. Ristic, J. Park, B. Q. Lv, Z. Wang, C. E. Matt, N. Xu, T. Shang, K. Conder, J. Mesot, S. Johnston, M. Shi and M. Radovic,
Phys. Rev. Lett. \textbf{117}, 037002 (2016).


    \bibitem{Franchini}
C. Franchini, G. Kresse, and R. Podloucky, Phys. Rev. Lett.  \textbf{102},
256402 (2009).

           \bibitem{Franchini2}
C. Franchini, A. Sanna, M. Marsman, and G. Kresse, Phys. Rev.
B \textbf{81}, 085213 (2010).

           \bibitem{Korotin}
D. Korotin, V. Kukolev, A. V. Kozhevnikov, D. Novoselov, and
V. I. Anisimov, J. Phys.: Condens. Matter  \textbf{24}, 415603 (2012).

 \bibitem{Nourafkan}
R. Nourafkan, F. Marsiglio, and G. Kotliar, Phys. Rev. Lett. \textbf{109},
017001 (2012).

           \bibitem{Yin}
Z. P. Yin, A. Kutepov, and G. Kotliar, Phys. Rev. X  \textbf{3}, 021011
(2013).

           \bibitem{Korotin2}
 D. M. Korotin, D. Novoselov, and V. I. Anisimov, J. Phys.: Condens. Matter \textbf{26}, 195602 (2014).

           \bibitem{Mattheiss83}
L. F. Mattheiss, Phys. Rev. B \textbf{28}, 6629 (1983).
 
    \bibitem{Mattheiss82}
L. F. Mattheiss and D. R. Hamann, Phys. Rev. B \textbf{26}, 2686 (1982).

       \bibitem{Foyevtsova}
  K. Foyevtsova, A. Khazraie, I. Elfimov and G. Sawatzky, Phys. Rev. B \textbf{91}, 121114(R) (2015).

    \bibitem{Ignatov}
  A. Ignatov, Nuclear Instruments and Methods in Physics Research A \textbf{448} 332(2000).

   \bibitem{Blaha}
   P. Blaha, K. Schwarz, G. K. H. Madsen, D. Kvasnicka, and J. Luitz, WIEN2K, An Augmented Plane Wave + Local Orbitals Program for Calculating Crystal Properties (2001).

      \bibitem{Perdew}
  J. Perdew, K. Burke and Y.Wang, Phys. Rev. B \textbf{54}, 16533

    \bibitem{Harrison2}
    S. Froyen and W. A. Harrison, Phys. Rev. B \textbf{20}, 2420
(1979).

  \bibitem{McMahan}
    A. K. McMahan and R. M. Martin, Phys. Rev. B \textbf{38}, 6650
(1988).

\bibitem{Mostofi08}
A. A. Mostofi, J. R. Yates, Y.-S. Lee, I. Souza,
D. Vanderbilt, and N. Marzari, Computer Physics
Communications {\bf 178}, 685
 (2008), ISSN 0010-4655.

\bibitem{Kunes10}
J. Kunes, R. Arita, P. Wissgott, A. Toschi, H. Ikeda, and
K. Held, Comp. Phys. Commun. {\bf 181}, 1888 (2010).

\bibitem{Sugai}
S. Sugai, S. Uchida, K. Kitazawa, S. Tanaka, and A. Katsui
Phys. Rev. Lett. {\bf 55}, 426 (1985).

\bibitem{Tajima}
S. Tajima, M. Yoshida, N. Koshizuka, H. Sato, and S. Uchida
Phys. Rev. B {\bf 46}, 1232 (1992).

\bibitem{Braden}
M. Braden, W. Reichardt, A. S. Ivanov, and A. Yu. Rumiantsev, Anamolous Dispersion of LO Phonon Branches in Ba$_{0.6}$K$_{0.4}$BiO$_{3}$, Europhys. Lett. B {\bf 34}, 531 (1996).


\bibitem{Hamada}
 N. Hamada, S. Massidda, A. J. Freeman, and J. Redinger, Phys.
Rev. B {\bf 40}, 4442 (1989).

\bibitem{Shirai}
M. Shirai, N. Suzuki, and K. Motizuki, J. Phys.: Condens. Matter
{\bf 2}, 3553 (1990).

 \bibitem{Liechtenstein}
  A. I. Liechtenstein, I. I. Mazin, C. O. Rodriguez, O. Jepsen, O. K. Andersen and M. Methfessel, Phys. Rev. B \textbf{44}, 5388 (1991). 

     \bibitem{Kunc}
 K. Kunc and R. Zeyher, Phys. Rev. B \textbf{49}, 12216 (1994).

      \bibitem{Meregalli}
 V. Meregalli and S. Y. Savrasov, Phys. Rev. B \textbf{57}, 14453 (1998).

      \bibitem{Bazhirov}
T. Bazhirov, S. Coh, S. G. Louie, and M. L. Cohen, Phys. Rev. B \textbf{88}, 224509 (2013). 

   

\end{thebibliography}
\end{document}